\documentclass[useAMS,usenatbib]{mn2e}

\usepackage{natbib}
\usepackage{aas_macros}
\usepackage{graphicx}
	\newcommand{\beq}{\begin{equation}}
	\newcommand{\eeq}{\end{equation}}
	\newcommand{\beqry}{\begin{eqnarray}}
	\newcommand{\eeqry}{\end{eqnarray}}
	\newcommand{\del}[2]{\frac{\partial #1}{\partial #2}}
\title[Assessment of Cowling approximation]{Assessment of Cowling approximation in computing ellipticity of a magnetized non-barotropic star}

\author[S'i. Yoshida]{Shin'ichirou Yoshida
\thanks{E-mail: yoshida@ea.c.u-tokyo.ac.jp}\\
Department of Earth Science and Astronomy,
Graduate School of Arts and Sciences, The University of Tokyo,\\
Komaba, Meguro-ku, Tokyo 153-8902, Japan
}

\begin{document}

\date{Accepted  Received ; in original form }

\pagerange{\pageref{firstpage}--\pageref{lastpage}} \pubyear{20xx}

\maketitle

\label{firstpage}

\begin{abstract}

A deformation of a neutron star due to its own magnetic field is an important issue
in gravitational wave astronomy, since a misaligned rotator with small ellipticity may emit
continuous gravitational wave that may be observed by ground-based detectors.
Recently Mastrano et al. \citep{Mastrano11, Mastrano13} evaluated deformations
induced by both poloidal and toroidal magnetic field in non-barotropic model stars
by neglecting the gravitational field perturbation (Cowling
approximation). Following their treatment in non-barotropic fluid and magnetic configurations, we 
here assess the effect of gravitational perturbation that they neglected. We show that the ellipticity
computed with gravitational perturbation is roughly twice as large as that obtained by
Cowling approximation. We should allow this amount of error in using the neat analytic
treatment proposed by them.

\end{abstract}

\section{Introduction}
A rotating neutron star with a non-axisymmetric deformation with respect to the rotational axis 
emits gravitational wave because of its time-dependent mass quadrupole. The quadrupole
scales as the ellipticity $\epsilon$ of the deformed star (see Eq.(\ref{eq: ellipticity defined})).
Therefore there is a strong astrophysical motivation to compute how much ellipticity we may expect 
in neutron stars.
One of the possibilities that have been proposed to allow misaligned deformation of rotating neutron stars
is the magnetic deformation of the stars. Radio pulsars are regarded as a misaligned rotator
with strong magnetic field (with typical surface fields $B\sim 10^8$ T ($10^{12}$ Gauss)). We now have
another class of neutron stars with super-strong field ($B\ge 10^9$ T), i.e.,'magnetars'.
Although the magnetic fields outside the stars are usually assumed to be dipolar, the configuration 
of magnetic field inside the stars is unknown. \citet{ChungMelatos2011b} gives observational
evidences of the dipolar field in some of the millisecond pulsars, while there are indications
that some of the neutron stars may have multipole magnetic field (see \citet{Mastrano13}).

The deformation of a neutron star due to magnetic
field strongly depends on its internal field configuration and we may have a clue to the internal
magnetic configuration when the gravitational waves from misaligned rotators are observed.

The studies on the issue of magnetic deformation of neutron stars have been 
mainly studied under the assumptions of barotropic stars
( general relativistic studies are found for instance in \citet{Bonazzola1996, Colaiuda2008, Ciolfi2010};
Newtonian counterparts are found in \citet{TomimuraEriguchi, YYE2006, Haskell2008, LanderJones2009, Fujisawa2012}).

Strictly speaking a neutron star is, however, a non-barotropic star which has a density stratification due to 
a gradient of chemical composition \citep{ReiseneggerGoldreich1992,Reisenegger2009}. In this case
the restrictions on a possible magnetic field structure inside may be rather relaxed compared to the
barotropic stars. This is because the gas pressure $p$ is not restricted by an one-parameter equation
of state of density $\rho$ such as $p=p(\rho)$ .

Under this physical picture, \citet{Mastrano11, Mastrano13} evaluated deformations
induced by both poloidal and toroidal magnetic field in non-barotropic model stars. They first prescribe
the functional form of magnetic field and compute the density perturbation induced by the Maxwell
stress. In these studies the perturbation of gravitational potential is neglected (Cowling approximation).
The former studies of deformations of neutron stars with crust suggest that Cowling approximation
is 
 According to the stellar perturbation
theory \citep{Unno_etal1989}, the perturbed gravitational potential for non-axisymmetric perturbations
becomes asymptotically zero with increasing number of azimuthal nodes. 
For axisymmetric perturbation of a lower order this is not the case, and we may
expect a large errors in neglecting gravitational perturbation. It is therefore to important to know how large
errors we may have in adopting Cowling approximation to compute a small deformation of a star by
magnetic stress.

In this paper we compare the ellipticity of magnetized non-barotropic star with and without gravitational
perturbation, by following the formulation in \citet{Mastrano11, Mastrano13}.

\section{Formulation}
We mainly follow the formulation made by \cite{Mastrano11}, except for the treatment
of density perturbation in the presence of magnetic field. Magnetic field is treated as a perturbation
to a non-magnetic spherical star with density and pressure stratification. The density profile of the non-magnetic star
is prescribed and the pressure profile is adjusted so that it balances the local gravitational acceleration.
Here we consider their "parabolic" density profile $\rho(r)$,
\beq
	\rho = \frac{15M_\star}{8\pi R_\star^3}(1-r^2),
	\label{eq: parabolic rho}
\eeq
where $M_\star, R_\star$ are the stellar mass and radius. The radial coordinate $r$ is normalized
by $R_\star$. The gravitational potential $\Phi$ consistent with the density profile is
\beq
	\Phi = \frac{GM_\star}{2R_\star}\left(\frac{5r^2}{2}-\frac{3r^4}{4}-\frac{15}{4}\right)
\eeq
and the stratified pressure distribution $p(r)$ balancing with the gravity is
\beq
	p = \frac{15M_\star^2}{16\pi R_\star^4}\left(1-\frac{5r^2}{2}+2r^4-\frac{r^6}{2}\right).
\eeq

Axisymmetric magnetic field is imposed on the non-magnetic star above. The magnetic field 
$\vec{B}$ is determined by two scalar functions $\alpha$ and $\beta$ as
\beq
	\vec{B} = B_0[\eta_{_P}\nabla\alpha\times\nabla\phi + \eta_{_T}\beta\nabla\phi],
	\label{eq: B-field}
\eeq
where $B_0$ is a constant and $\beta$ is a functional of $\alpha$. $\eta_{_P}$ and $\eta_{_T}$ are parameters determining
the poloidal and toroidal magnetic field amplitude. We adopt the analytic profile of $\alpha$ 
as in \cite{Mastrano11} and their functional form of $\beta$
\beq
	\alpha = \frac{35}{8}\left(r^2-\frac{6r^4}{5}+\frac{3r^6}{7}\right),
\label{eq: magnetic-alpha}
\eeq
and
\beq
	\beta = \left\{\begin{array}{cc}
			(\alpha-1)^2 & (\alpha\ge 1)\\
			0 & (\alpha< 1).
		\end{array}\right.
\label{eq: magnetic-beta}
\eeq
Lorentz force by the magnetic field modifies the density and pressure distribution inside the star. The magnetic
stress is treated as a small perturbation and we consider the linear perturbation $\delta\rho, \delta p$ and
the perturbation of gravitational potential $\delta\Phi$. The force balance including magnetic stress is
\beq
	\frac{1}{\mu_0}(\nabla\times\vec{B})\times\vec{B} = \nabla\delta p + \delta\rho\nabla\Phi + \rho\nabla\delta\Phi,
\label{eq: forcebalance}
\eeq
where the last term on the right hand side is neglected in \cite{Mastrano11}. We here adopt SI unit. 
By integrating $\theta$-component of Eq.(\ref{eq: forcebalance}) we obtain the following relation,
\beqry
	\delta p + \rho\delta\Phi &=& -\frac{B_0^2\alpha}{\mu_0r^2\sin^2\theta}
	\left[\eta_{_P}^2\hat{\triangle}\alpha \right.\nonumber\\ 
	&& \left.+ 2\eta_{_T}^2 \left(\frac{\alpha^3}{3}-\frac{3\alpha^2}{2}+3\alpha-\ln\alpha - \frac{11}{6}\right)
	\right],
\label{eq: forcebalance-theta}
\eeqry
where the operator $\hat{\triangle} = \del{{^2}}{r^2} + \frac{\sin\theta}{r^2}\del{ }{\theta}\left(\frac{1}{\sin\theta}\del{ }{\theta}\right)$
is introduced. Here we omit a monopole term, as is done in \cite{Mastrano11}, since it amounts to rescaling the spherical
background star. Then $r$-component of Eq.(\ref{eq: forcebalance}) is written as,
\beqry
	\frac{d\Phi}{dr}\delta\rho - \frac{d\rho}{dr}\delta\Phi &=&
	\frac{B_0^2}{\mu_0}\eta_{_P}^2\alpha\del{ }{r}\left(\frac{\hat{\triangle}\alpha}{r^2\sin^2\theta}\right)\nonumber\\
	&&+ \frac{B_0^2}{\mu_0}\cdot 2\eta_{_T}^2
		\left(
			\frac{\alpha^3}{3}-\frac{3\alpha^2}{2} \right.\nonumber\\ && \left.+3\alpha-\ln\alpha - \frac{11}{6}
		\right)\del{ }{r}\left(\frac{\alpha}{r^2\sin^2\theta}\right).
\label{eq: forcebalance-r}
\eeqry
The right hand side is a fixed function of $r,\theta$ when $\alpha$ and $\beta$ are prescribed. 
It should be noticed that the terms in (\ref{eq: forcebalance-theta}) and (\ref{eq: forcebalance-r}) having
the factor $\eta_{_T}^2$ vanishes when $\alpha <1$. It ensures that the toroidal component of magnetic
field is enclosed in the star and no electric current exists in vacuum outside.
When Cowling
approximation is adopted as in \cite{Mastrano11}, the second term on the left hand side is neglected 
and $\delta\rho$ is evaluated analytically. 
We here retain the term and compare the results with those by Cowling approximation. 
$\delta\Phi$ in the second term is a functional of $\delta\rho$ and written as
\beq
	\delta\Phi(r,\theta) = 4\pi G\int dV'~ \delta\rho(r',\theta) {\cal G}(r',\theta'|r,\theta)
\eeq
by using an appropriate Green's function ${\cal G}$ for Poisson's equation.
We solve Eq.(\ref{eq: forcebalance-r}) for $\delta\rho$ on a finite-difference grid point by discretizing the integration.

Ellipticity $\epsilon$ that characterizes the stellar deformation due to magnetic field is defined by
\beqry
	\epsilon &=& \frac{I_{zz}-I_{xx}}{I_0}\nonumber\\
	&=& \pi R_\star^5 I_0^{-1} \int_0^1r^2 dr
		\int_0^\pi \sin\theta d\theta~ \delta\rho ~ r^2 (1-3\cos^2\theta),
\label{eq: ellipticity defined}
\eeqry
where $I_0$ is the moment of inertia of the non-magnetic spherical star.

The ratio of magnetic field strength in poloidal and toroidal component is measured by the parameter $\Lambda$
which is defined as the ratio of the magnetic energy of poloidal field to the total magnetic energy 
(the domain of the integrations also includes the vacuum region around the stars),
\beq
	\Lambda = \frac{\int dV \vec{B}_p^2}{\int dV \vec{B}^2}
	= \frac{\eta_{_P}^2}{\eta_{_P}^2 + q\eta_{_T}^2}.
\eeq
Here $\vec{B}_p$ is the first term of Eq.(\ref{eq: B-field}).
For the prescribed magnetic field (\ref{eq: magnetic-alpha}) and (\ref{eq: magnetic-beta}), we have $q=1.95\times 10^{-6}$.

\section{Results}
\begin{figure}
\includegraphics[scale=1.0]{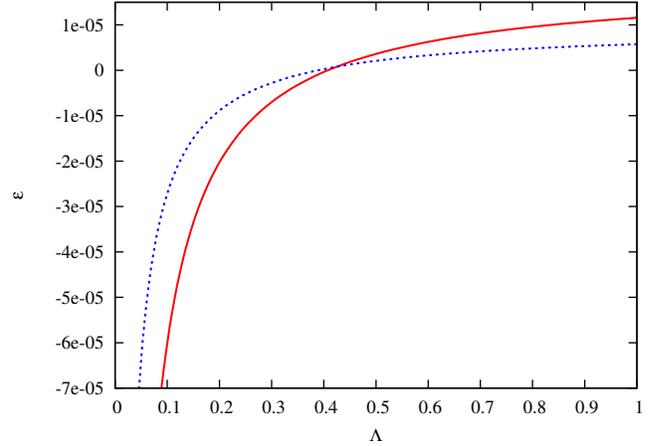}%
\caption{\label{fig: parabolic case}
Ellipticity of a stellar model induced by magnetic field (Eq.(\ref{eq: B-field})) is plotted as a function of $\Lambda$
which measures the relative strength of poloidal magnetic field to the total field. The solid line
takes into account gravitational perturbation, while the dotted line is for Cowling approximation.
The non-magnetic stellar model has a parabolic density profile as Eq.(\ref{eq: parabolic rho}).
We fix $B_0^2/(\mu_0 p_c)$ to be $10^{-7}$ and $\eta_P$ to be unity.
}
\end{figure}

We first compare the ellipticity computed with and without gravitational perturbation by
fixing other model details as in \cite{Mastrano11}. In Fig.\ref{fig: parabolic case} we plot the ellipticity $\epsilon$ 
as a function of $\Lambda$ which measures the relative energy of poloidal magnetic field to the total one. 
The solid line corresponds to the full perturbation when the perturbation of gravitational potential is taken into account.
The dashed line is for Cowling approximation as in \cite{Mastrano11}. For this parabolic density profile, the ellipticity 
of the full perturbation is nearly twice as large as that of Cowling approximation. 
The fitting formula of $\epsilon$ is
	\beqry
		\epsilon = 1.3\times 10^{-5} \left(\frac{B_{\rm surface}}{5\times 10^{10}{\rm T}}\right)^2
		\left(\frac{M_\star}{1.4M_\odot}\right)^{-2}
		\left(\frac{R_\star}{10^4{\rm m}}\right)^4\nonumber\\
		\times\left(1-\frac{0.41}{\Lambda}\right),
	\eeqry
where $B_{\rm surface}\equiv\eta_PB_0$.

\begin{figure}
\includegraphics[scale=1.0]{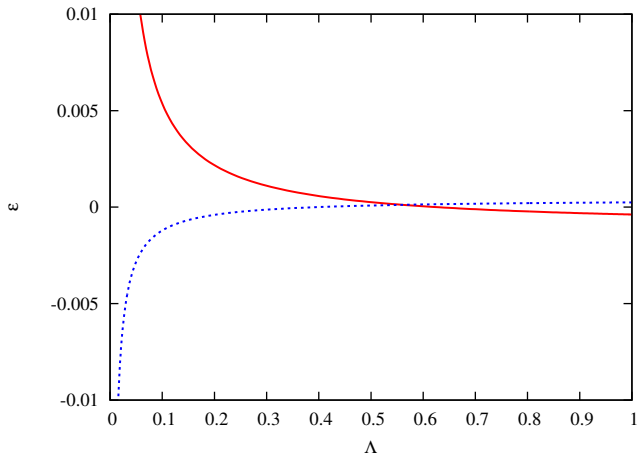}%
\caption{\label{fig: off-centred case}
The same as Fig.\ref{fig: parabolic case} except that the non-magnetic stellar model has an
off-centred density maximum.
}
\end{figure}

We also compute the ellipticity for
the polytropic case with index $N=1$ which is studied in \cite{Mastrano11}. Compared to the parabolic case, the 
density profile of the polytrope corresponds to that of a slightly softer equation of state, which shows more concentration
near the origin. Overall the density profile is similar to the parabolic one and we confirm that the relative error in 
Cowling approximation is almost the same as in the parabolic case. 
$\epsilon$ for this case is fitted as,
	\beqry
		\epsilon = 1.2\times 10^{-5} \left(\frac{B_{\rm surface}}{5\times 10^{10}{\rm T}}\right)^2
		\left(\frac{M_\star}{1.4M_\odot}\right)^{-2}
		\left(\frac{R_\star}{10^4{\rm m}}\right)^4\nonumber\\
		\times\left(1-\frac{0.40}{\Lambda}\right).
	\eeqry
This is, however, only for these particular
choices of density and magnetic profile. To see what happens for a different situation, we adopt an extreme case of 
density profile. A density profile with off-centred maximum is defined as
\beq
	\rho = \frac{15M_\star}{8\pi R_\star^3}\left(r^2 - r^3\right)
\eeq
with a corresponding gravitational potential 
\beq
	\Phi = \frac{GM_\star}{2R_\star}\left(\frac{r^4}{20}-\frac{r^6}{42}-\frac{851}{420}\right).
\eeq
Although this profile is far from realistic, it gives an example showing that the density profile of the non-magnetic
background state is as important to determine $\epsilon$ as the magnetic field distribution. In Fig.\ref{fig: off-centred case}
we plot the ellipticity as a function of $\Lambda$ parameter. The magnetic field distribution is the same as in the
parabolic density profile case above. We are here interested in the qualitative behaviour and do not care the 
size of the ellipticity. The ellipticity in Cowling approximation behaves rather differently from that in full perturbation.
For a larger poloidal fraction of magnetic field, the full theory predicts prolate deformation of a star
while the Cowling approximation produces an oblately deformed star. As the toroidal fraction is increased
( $\Lambda\to 0$), this tendency is reversed. This extreme case shows that we need to take into account gravitational
perturbation to compute securely the ellipticity of a star.

%
%
\begin{figure}
\includegraphics[scale=1.0]{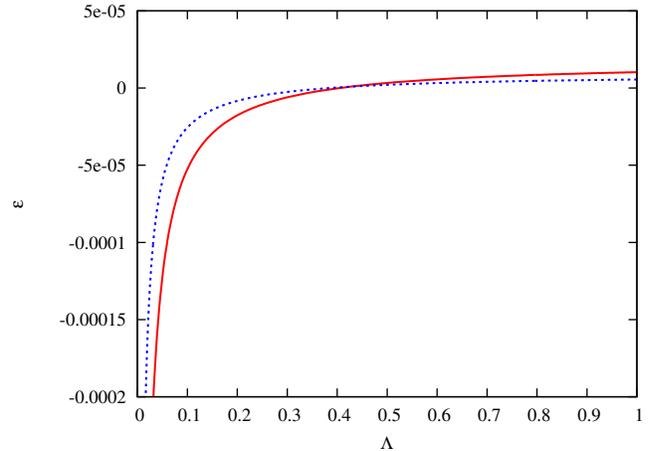}%
\caption{\label{fig: polyN1 case}
The same as Fig.\ref{fig: parabolic case} except that the non-magnetic stellar model has
a density profile of $N=1$ polytrope.
}
\end{figure}

%

\section{Summary and Comments}
\citet{Mastrano11, Mastrano13} computed the ellipticity induced by the Maxwell stress of density-stratified stars
with magnetic field. It is assumed that the deformation is so small that
it is treated as a linear perturbation to the non-bartotropic spherical equilibrium. With Cowling approximation
in which the Eulerian perturbation of gravitational potential is neglected, they obtain a neat analytic
formula for the ellipticity assuming magnetic field to have a prescribed analytic distribution.
We assessed the reliability of Cowling approximation in their model by taking the perturbed gravitational
potential into account. As is expected from the stellar perturbation theory, Cowling approximation gives rise to
a large error in density distribution and the ellipticity depending on the spherical background star as well
as on the magnetic distribution. For a comparatively realistic case, however, the deviation of the ellipticity 
due to Cowling approximation from the actual value may be smaller than an order of magnitude.
The result is consistent with the former studies \citep{Ushomirsky_etal2000,CutlerUshomirskyLink2003,HaskellJonesAndersson2006}
in which the effects of Cowling approximation on the ellipticity due to crust mountains of neutron stars are estimated.
Although the magnetic field instead of the crust shear modulus deforms the stars here, 
the results obtained are consistent with those in the former studies, i.e., neglecting the gravitational perturbation
may suffer a few hundred per cent of errors in the ellipticity for astrophysically relevant models.
On the other hand, by comparing the full and the Cowling treatment for non-trivial density distribution
we see that neglecting gravitational potential may lead to a rather erroneous result. It may be safe to
take into account the gravitational perturbation in computing magnetic deformation of a star.

Finally an extension of the formulations by Tokyo group \citep{TomimuraEriguchi, YYE2006, Fujisawa2012} 
to include non-barotropic effect may be useful to compute consistently the effect of magnetic field
and gravity on the ellipticity of magnetized stars.

\section*{Acknowledgments}
The author thanks the anonymous reviewer for useful comments.

\bibliographystyle{mn}

\bibliography{mn-jour,pap}

\label{lastpage}

\end{document}